\documentclass{caosp305}

\usepackage{graphicx}

\usepackage{natbib}
\articleNo{123}
\pubyear{2017}
\volume{35}
\volnumber{3}
\firstpage{1}
\received{December 2017}
\accepted{December 2017}

\def\BibTeX{{\rm B\kern-.05em{\sc i\kern-.025em b}\kern-.08em
             T\kern-.1667em\lower.7ex\hbox{E}\kern-.125emX}}

\begin{document}

\hauthor{P.\,Padoan}

\title{The magnetic field of molecular clouds}

\author{
        P.\,Padoan \inst{1,} \inst{2} 
       }

\institute{
           Institut de Ci\`{e}ncies del Cosmos, Universitat de Barcelona, IEEC-UB, Mart\'{i} i Franqu\`{e}s 1, E08028 Barcelona, Spain, \email{ppadoan@icc.ub.edu}
         \and 
           ICREA, Pg. Llu\'{i}s Companys 23, 08010 Barcelona, Spain 
          }

\date{December 11, 2017}

\maketitle

\begin{abstract}

The magnetic field of molecular clouds (MCs) plays an important role in the process of star formation: it 
determins the statistical properties of supersonic turbulence that controls the fragmentation of MCs, 
controls the angular momentum transport during the protostellar collapse, and affects the stability of
circumstellar disks. In this work, we focus on the problem of the determination of the magnetic field 
strength. We review the idea that the MC turbulence is super-Alfv\'{e}nic, and we argue that MCs are
bound to be born super-Alfv\'{e}nic. We show that this scenario is supported by results from a recent 
simulation of supernova-driven turbulence on a scale of 250\,pc, where the turbulent cascade is resolved 
on a wide range of scales, including the interior of MCs. 

\end{abstract}

\section{The magnetic-field strength in MCs}

The idea that MCs are magnetically supported against their gravitational collapse was reviewed 
in \citet{Shu+87}. In that scenario, the observed random velocities correspond to MHD waves, or perturbations
of a strong mean field. Gravitationally bound prestellar cores are initially subcritical and contract because of 
ambipolar drift until they become supercritical and collapse. 
\citet{Padoan+Nordlund97MHD,Padoan+Nordlund99MHD} proposed an alternative scenario where
the mean magnetic field in MCs is weak and the observed turbulence is super-Alfv\'{e}nic. By comparing results 
of two simulations, one with a weak field and the other with a strong field, with observational data, they showed that 
the super-Alfv\'{e}nic case reproduced the observations better. Further results in support of the super-Alfv\'{e}nic scenario 
were later presented in \citet{Padoan+04power} and, more recently, by \citet{Lunttila+08,Lunttila+09} based on 
simulated Zeeman measurements.

In this review, we first address the super-Alfv\'{e}nic nature of the turbulence in MCs in the context 
of their formation process. To support our scenario, we present results of a large-scale (250\,pc) MHD simulation of
interstellar medium (ISM) turbulence driven by supernova (SN) explosions. We then show that MC turbulence is super-Alfv\'{e}nic
also with respect to their rms magnetic field, amplified by the turbulence. Finally, we briefly summarize observational results in 
favor of this picture.

\section{MCs are born super-Alfv\'{e}nic}

Different processes may contribute to the formation of MCs, and a full treatment of this problem should
also address the mechanisms of conversion between atomic and molecular gas. For a general picture,
we assume that MCs are the result of large-scale compressions of the warm interstellar medium (WISM), 
driven by the evolution of SN remnants. When such compressions reach the pressure threshold of the thermal 
instability, the compressed gas rapidly cools and compresses further to a characteristic density of MCs.  
We can characterize the large-scale turbulence of the WISM by its rms sonic and Alfv\'{e}nic Mach 
numbers, $M_{\rm s}$ and $M_{\rm A}$. It is generally believed that the large-scale turbulence in the WISM
is transonic and trans-Alfv\'{e}nic, meaning $M_{\rm s}\sim 1$ and $M_{\rm A}\sim 1$. 

\begin{figure}[t]
  \includegraphics[height=.25\textheight]{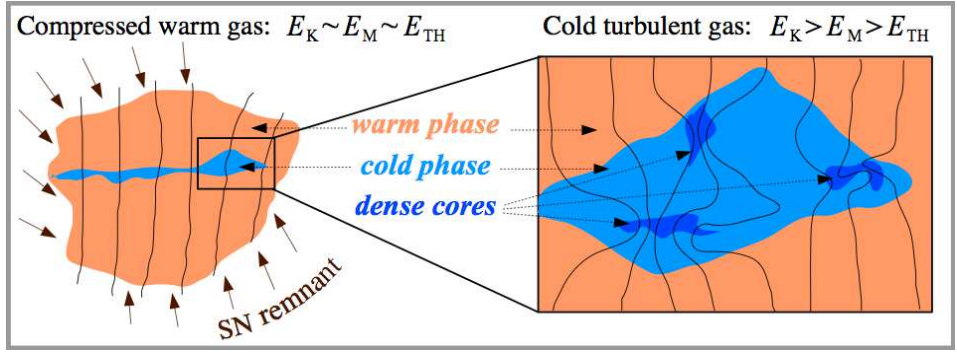}
  \caption{Schematic scenario of the formation of super-Alfv\'{e}nic MCs.}
  \label{f1}
\end{figure}

Because of the transonic nature of the WISM turbulence, the large-scale velocity field can occasionally
cause compressions strong enough to bring large regions above the thermal instability threshold. As the 
gas is further compressed thanks to its cooling, the magnetic field cannot be compressed because of 
the initially trans-Alfv\'{e}nic nature of the flow, so the initial compression is forced to be primarily along the 
magnetic field direction. Assuming that turbulent velocities are not significantly decreased, the characteristic 
increase in density, $\rho_{\rm cold}\sim 100\, \rho_{\rm warm}$, results in a comparable increase in the turbulent 
kinetic energy, $E_{\rm K,cold}\sim 100\, E_{\rm K,warm}$, or a corresponding drop in the rms Alfv\'{e}n velocity, 
$V_{\rm a,cold}\sim V_{\rm a,warm}/10$. As a consequence, the turbulence in the rapidly cooling gas must be 
initially super-Alfv\'{e}nic with respect to the mean magnetic field \citep{Nordlund+Padoan03,Padoan+10_Como}. 
Compression and stretching in this super-Alfv\'{e}nic flow can then locally amplify the magnetic field, without affecting 
the mean field. Because of the reduced temperature, the turbulence in the cold gas is also supersonic, so dense cores 
with enhanced magnetic field strength are naturally formed by shocks in the turbulent flow. This sequence 
of events is schematically depicted in Fig.\,\ref{f1}. 

\section{Supernova-driven turbulence}

The above scenario of MC formation is supported by a recent SN-driven MHD 
simulation \citep{Padoan+16SN_I,Pan+16,Padoan+16SN_III,Padoan+17sfr} that models a 250-pc ISM region, 
large enough to address the formation and evolution and MCs. The MHD equations are solved with the Ramses 
AMR code \citep{Teyssier02,Fromang+06,Teyssier07} within a cubic region of size $L_{\rm box}=250$ pc, 
total mass $M_{\rm box}=1.9\times 10^6$ $M_{\odot}$, and periodic boundary conditions. The mean density 
is $n_{\rm H,0}=5$ cm$^{-3}$ and the mean magnetic field $B_0=4.6$ $\mu$G. The rms magnetic field amplified 
by the turbulence is 7.2 $\mu$G. 

The only driving force is from SN feedback. SNe are randomly distributed in space and time during the first period of the simulation without
self-gravity, while they are later determined by the position and age of the massive sink particles formed when self-gravity is included. In the 
initial phase without gravity, the minimum cell size is $dx=0.24$ pc until $t=45$ Myr. It is then decreased to $dx=0.03$ pc, using a root-grid 
of $512^3$ cells and four AMR levels, during an additional period of 10.5 Myr without self-gravity. Finally, at $t=55.5$ Myr, gravity is introduced 
and the minimum cell size is further reduced to $dx=0.0076$ pc by adding two more AMR levels. 

To follow the collapse of prestellar cores, sink particles are created in cells where the gas density is larger than $10^6$ cm$^{-3}$ if a
number of conditions are met \citep[see][]{Haugboelle+17imf}. When a sink particle of mass larger than 7.5 M$_{\odot}$ has an age equal to the 
corresponding stellar lifetime for that mass, a sphere of $10^{51}$ erg of thermal energy is injected at the location of the sink 
particle to simulate the SN explosion, as described in detail in \citet{Padoan+16SN_I}. We refer to this driving method as \emph{real SNe}, as 
it provides a SN feedback that is fully consistent with the star-formation rate (SFR), the stellar initial mass function, and the ages and positions 
of the individual stars whose formation is resolved in the simulation. The simulation has so far been run for approximately 30 Myr with self-gravity, 
star formation and \emph{real SNe}, generating over 7000 stars and hundreds of MCs. The SFR in the MCs has realistic values, while the global 
SFR corresponds to a mean gas-depletion time in the computational volume of almost 1 Gyr, also realistic for a 250-pc scale \citep{Padoan+17sfr}.

\section{The mean and rms magnetic-field strength of MCs}

\begin{figure}[t]
\includegraphics[width=6.cm,clip=]{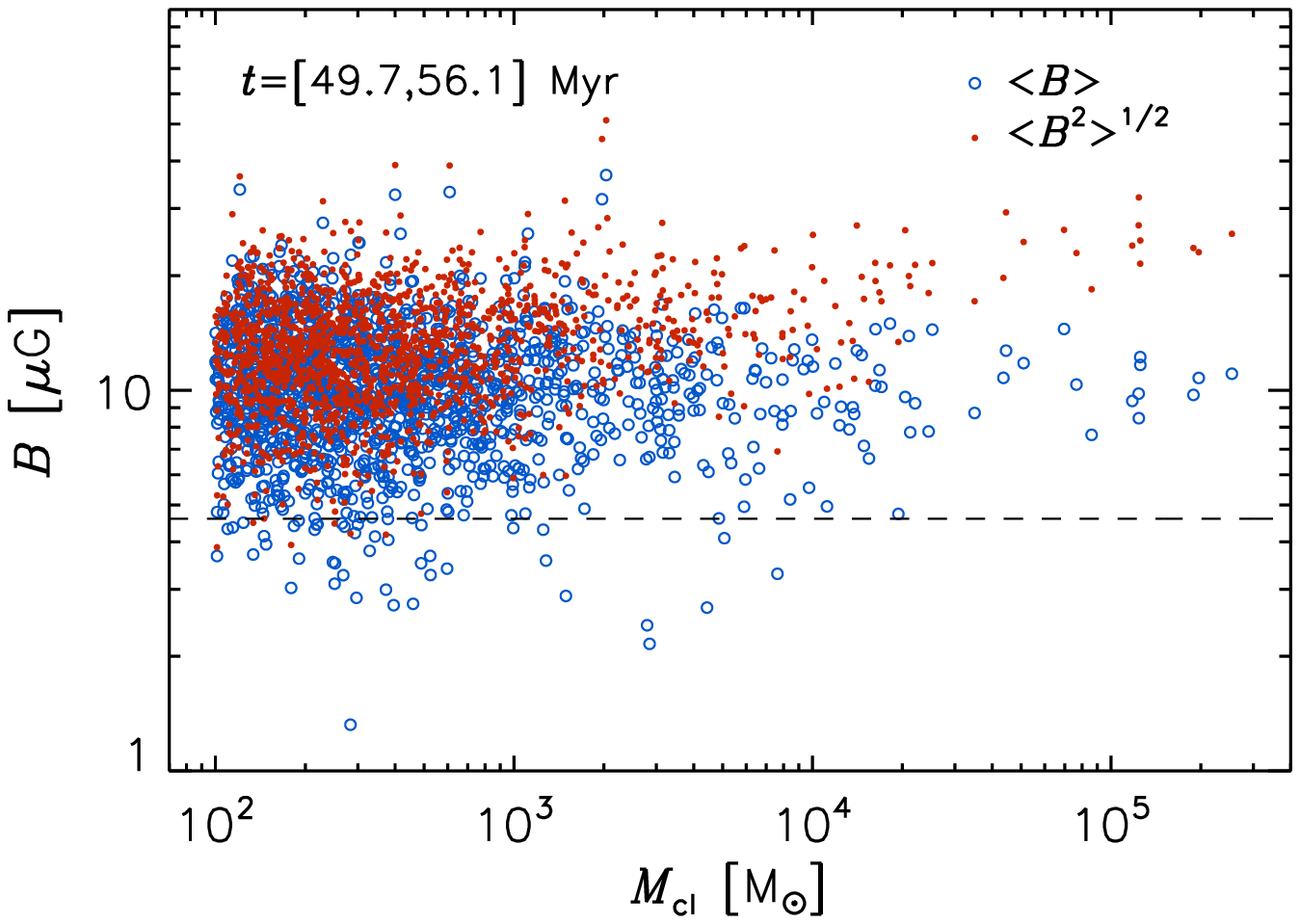}
\includegraphics[width=6.cm,clip=]{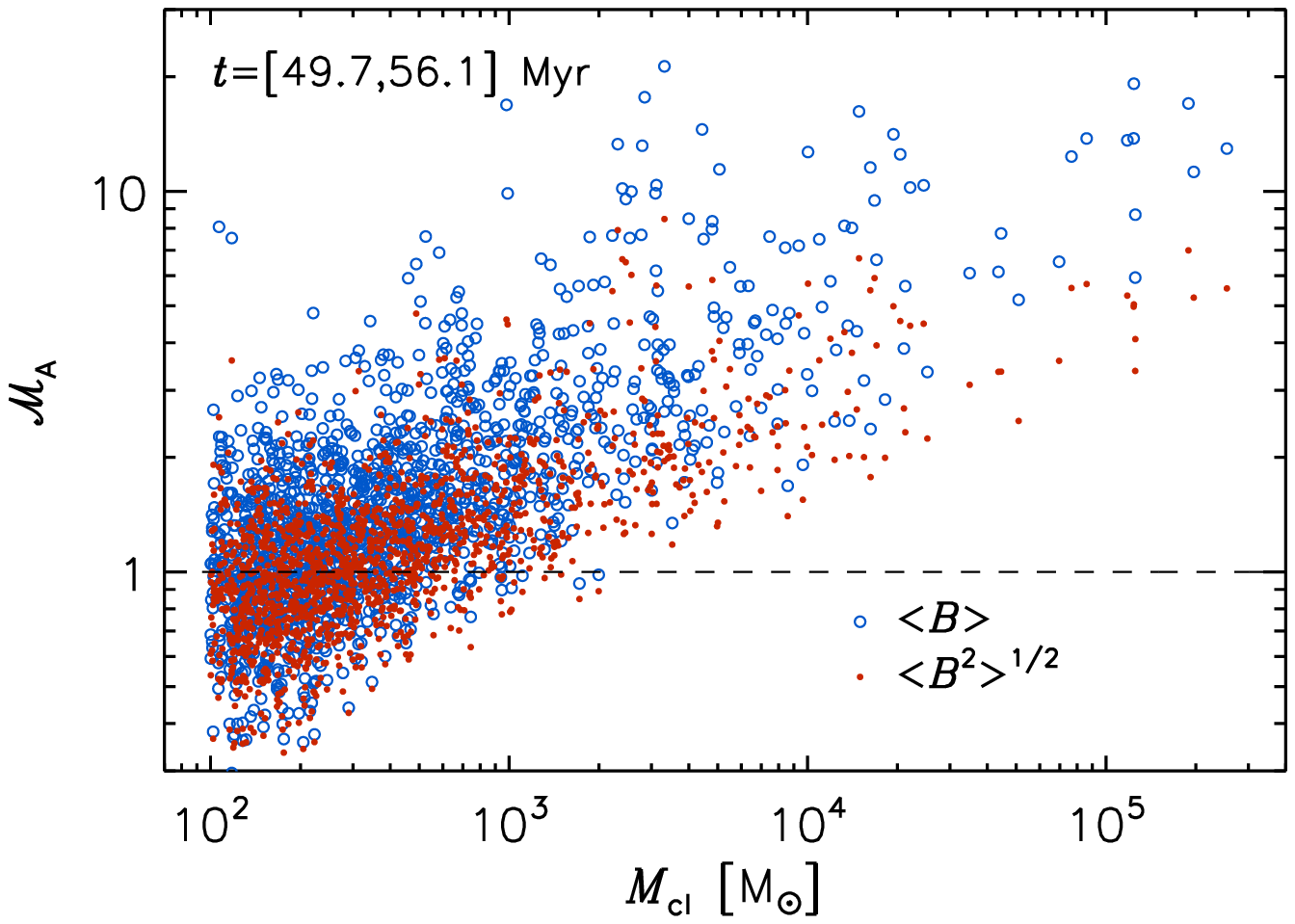}
\caption[]{{\it Left}: Magnetic field strength versus cloud mass for a sample of 1,547 MCs selected from our SN-driven simulation. 
The dashed line shows the mean magnetic field averaged over the whole computation volume (also the initial mean field). Empty 
circles correspond to the mean value of the magnetic field of all tracer particles in each cloud, while filled circles give the rms value. 
{\it Right}: Alfv\'{e}nic rms Mach number versus cloud mass for the same clouds as in the left panel, computed with the cloud mean 
magnetic field (empty circles), or the cloud rms magnetic field (filled cirlces).}
\label{f2}
\end{figure}

The simulation adopts a mean magnetic-field strength consistent with the Galactic one, so the magnetic field inside clouds selected 
from the simulation should be comparable to that in real MCs. To investigate the role of the magnetic field in individual MCs, in \citet{Padoan+16SN_I}
we consider a catalog of 1,547 clouds selected from several snapshots of the simulation, based on a density threshold of $n_{\rm H,min}=200$ cm$^{-3}$. 
The mean and rms magnetic field of each cloud is computed using the values sampled by tracer particles embedded in the simulation, 
$\langle B \rangle = \Sigma_{\rm i}B_{\rm i}/N$ and $\langle B^2 \rangle^{1/2} = (\Sigma_{\rm i}B^2_{\rm i}/N)^{1/2}$, where $B_{\rm i}$ is the magnetic 
field strength sampled by the particle i in a given cloud, and $N$ is the total number of particles in that cloud. These magnetic field values are plotted 
versus the cloud mass in the left panel of Fig.\,\ref{f2}, where the horizontal dashed line represents the mean magnetic field in the computational volume, 
$B_0=4.6$ $\mu$G. 

The mean field in the clouds (empty circles) is approximately 10 $\mu$G on average, only twice larger than the large-scale magnetic-field strength, 
$B_0$, and independent of cloud mass. We have verified that the mean magnetic-field strength of the clouds is also independent of their mean gas density. 
The relatively small increase of the cloud mean magnetic field relative to $B_0$ and its independence of gas density are characteristic of trans-Alfv\'{e}nic 
supersonic turbulence \citep{Padoan+Nordlund97MHD,Padoan+Nordlund99MHD}, and illustrates that MCs must be formed by compressive motions primarily 
along magnetic field lines, due to the non-negligible magnetic pressure prior to the compression and cooling of the low-density gas, as discussed above. 

Because in super-Alfv\'{e}nic  turbulence the magnetic field is amplified by compressions, as shown by a positive $B-n$ correlation 
\citep{Padoan+Nordlund97MHD,Padoan+Nordlund99MHD}, our scenario also predicts the formation of dense cores, formed by shocks within MCs 
\citep{Padoan+2001cores}, with an enhanced magnetic-field strength. The small-scale enhancement of the magnetic field within MCs is illustrated 
in the left panel of Fig.\,\ref{f2}, where the values of the rms field in the clouds (filled circles) is approximately a factor of two larger than the mean field.
The rms field increases slightly with cloud mass, due to the correlation between rms velocity and cloud mass: Larger, more massive 
clouds have larger rms velocity than smaller clouds, on average, so the rms field is amplified to a larger value.

As a more direct demonstration of the super-Alfv\'{e}nic nature of MC turbulence, the right panel of Fig.\,\ref{f2} shows the cloud rms Alfv\'{e}nic 
Mach number versus the cloud mass. The Mach number is computed as the ratio of the cloud rms velocity and the cloud Alfv\'{e}n velocity, where 
the latter is computed either with the mean magnetic field (empty circles) or with the rms magnetic field (filled circles), and using the mean density 
sampled by the tracer particles. Nearly all clouds with mass larger than $10^3$ M$_{\odot}$ are super-Alfv\'{e}nic, even considering their amplified 
field strength. For the 41 MCs with masses larger than $10^4$ M$_{\odot}$, the average Alfv\'{e}nic Mach number is 8.3 with respect to the mean field, 
and 3.9 with respect to the rms field. 

The inability of supersonic turbulence to amplify the magnetic field to equipartition with the kinetic energy, as in the MCs of
our simulation, was already established with idealized simulations of randomly-driven MHD turbulence. Simulations
by \citet{Haugen+04} suggested that the growth rate of the turbulent dynamo in the supersonic regime may be significantly 
reduced compared to the incompressible case. Later on, simulations with rms sonic Mach number ${\mathcal M}_{\rm s}\approx 10$ 
and different values of the rms Alfv\'{e}nic Mach number, ${\mathcal M}_{\rm A}\approx 30$, 10, and 3 
\citep{Kritsuk+09a,Kritsuk+09b,Padoan+10_Como}, also showed that the saturated rms magnetic field is a function of the
mean field, and it is consistent with amplification by compression, with no contribution from a turbulent dynamo 
\citep[see eq. (20) in][]{Padoan+Nordlund11sfr}. As firmly established in the comprehensive parameter study by
\citet{Federrath+11}, a supersonic turbulent dynamo plays a role in the field amplification only when the mean magnetic 
field is very small, as the saturated value of the magnetic energy is only a few percent of the kinetic energy. 

\begin{figure}[t]
\centering
\includegraphics[width=9.cm,clip=]{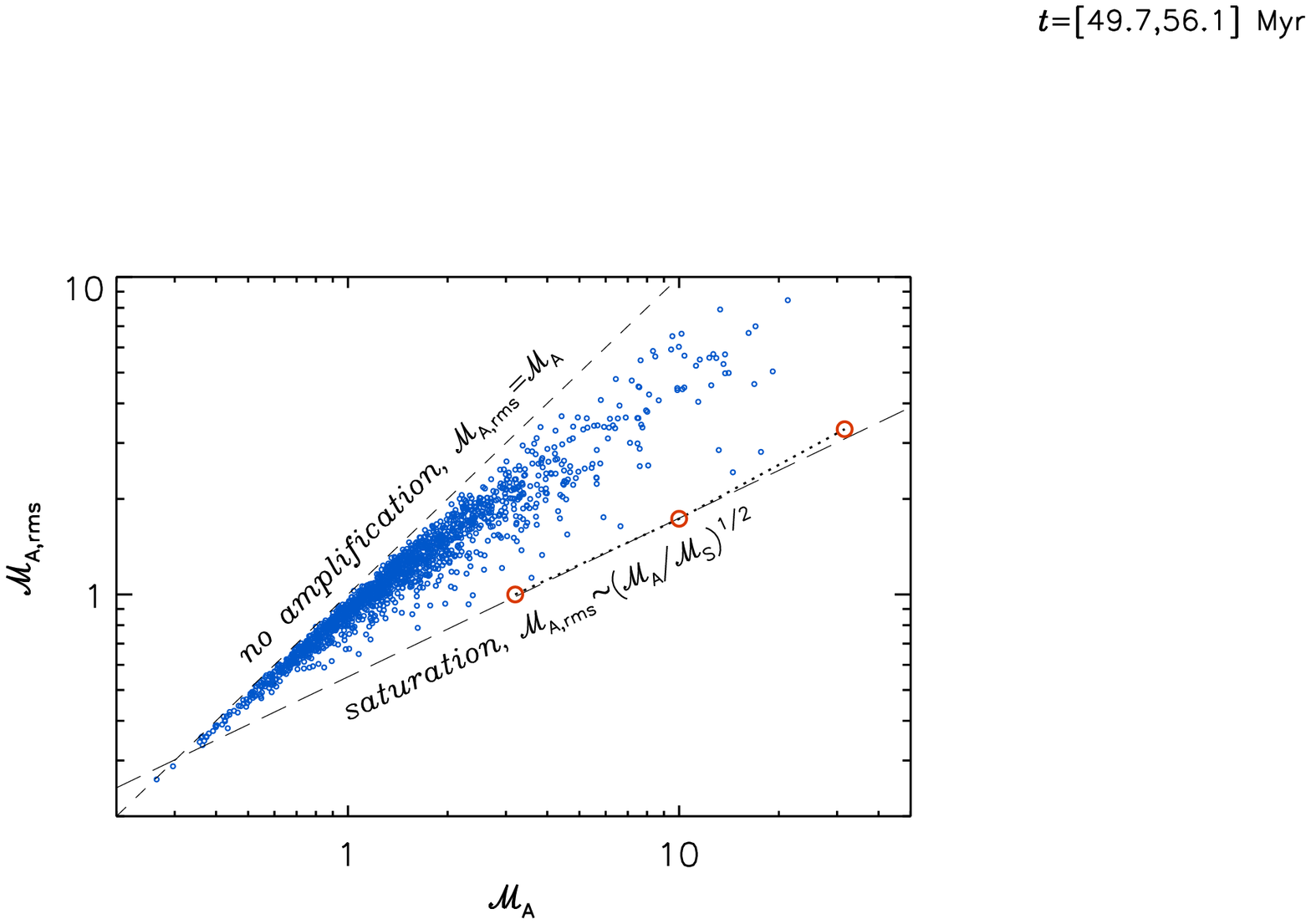}
\caption[]{The rms Alfv\'{e}nic Mach number computed with the rms magnetic field, ${\mathcal M}_{\rm A,rms}$, versus the Alfv\'{e}nic 
Mach number based on the mean magnetic field, ${\mathcal M}_{\rm A}$, for the same MCs selected from the SN-driven simulation
as in fig.\,\ref{f2}. The long-dashed line shows the saturation value from eq. (1), while the open circles indicate the saturation values 
from randomly driven simulations \citep{Kritsuk+09a,Kritsuk+09b,Padoan+10_Como}.}
\label{f3}
\end{figure}

Equation (20) in \citet{Padoan+Nordlund11sfr} corresponds to the following relation between the rms Alfv\'{e}nic Mach number
computed with the rms magnetic field, ${\mathcal M}_{\rm A,rms}$, and the ratio of Alfv\'{e}nic to sonic Mach numbers based 
on the mean magnetic field, ${\mathcal M}_{\rm A}$ and ${\mathcal M}_{\rm s}$ respectively:
\begin{equation}
{\mathcal M}_{\rm A,rms}\sim ({\mathcal M}_{\rm A}/{\mathcal M}_{\rm s})^{1/2}.
\end{equation}
This predicted saturation level is shown by the long-dashed line in fig.\,\ref{f3}. One can see that it defines the lower envelope of the
scatter plot of ${\mathcal M}_{\rm A,rms}$ versus ${\mathcal M}_{\rm A}$ for the MCs selected from our SN-driven simulation.
Many clouds are found well above the saturation level, which may be due to their relatively young age, or to an insufficient numerical
resolution in the case of the smallest clouds.

\section{Comparison with Observations}

The super-Alfv\'{e}nic nature of the turbulence in the clouds from our simulation is consistent with the observational evidence. Based on the comparison
between simulations of MHD turbulence and MC observations, \citet{Padoan+Nordlund97MHD,Padoan+Nordlund99MHD} suggested that MC turbulence
was better characterized by supersonic turbulent flows with ${\mathcal M}_{A}\gg 1$ than flows with  ${\mathcal M}_{A}\approx 1$. This result was later 
confirmed with the aid of synthetic observations \citep{Padoan+04power} and synthetic Zeeman splitting measurements \citep{Lunttila+08}. It was shown 
that a super-Alfv\'{e}nic turbulence simulation with the characteristic size, density, and velocity dispersion of star-forming regions could produce dense cores 
with the same relation between magnetic-field strength and column density as observed cores. \citet{Lunttila+08} also computed the relative mass-to-flux 
ratio ${\mathcal R}_{\mu}$, defined as the mass-to-flux ratio of a core divided by that of its envelope, as proposed by \citet{Crutcher+09}. They found a large 
scatter in the value of ${\mathcal R}_{\mu}$, and an average value of ${\mathcal R}_{\mu}<1$, in contrast to the ambipolar-drift model of core formation, where 
the mean magnetic field is stronger and only ${\mathcal R}_{\mu}>1$ is allowed. \citet{Crutcher+09} confirmed that ${\mathcal R}_{\mu}<1$ in observed cores, 
as predicted by \citet{Lunttila+08}.

In a separate work, \citet{Lunttila+09} used simulated OH Zeeman measurements to compute the mass-to-flux ratio relative to the critical one, $\lambda$, 
and the ratio of turbulent to magnetic energies, $\beta_{\rm turb}$, in molecular cores selected from a super-Alfv\'{e}nic simulation. They found both mean 
values and scatter of $\lambda$ and $\beta_{\rm turb}$ in good agreement with the observational results of \citet{Troland+Crutcher08}. In our super-Alfv\'{e}nic 
scenario, the scatter originates partly from intrinsic variations of the magnetic field strength from core to core, which are not expected in the traditional picture of 
MCs were the mean magnetic field is strong \citep{Shu+87}.

Taking advantage of the anisotropy of MHD turbulence, \citet{Heyer+Brunt12} demonstrated that the densest regions of the Taurus MC complex are characterized 
by super-Alfv\'{e}nic turbulence, while in low density regions the motions are sub or trans-Alfv\'{e}nic, also consistent with the picture from our simulation, where
MCs are formed by large-scale trans-Alfv\'{e}nic turbulence, and thus fed preferentially by motions along magnetic field lines, as discussed above
\citep{Nordlund+Padoan03,Padoan+10_Como}. Recent dust polarization measurements have shown that the magnetic field has a relatively constant orientation
in low-density regions surrounding MCs, with the mean direction mainly parallel to low-density filaments, while the field becomes predominantly perpendicular 
to the direction of dense filaments in regions of larger column density \citep{Soler+17}. This change is suggestive of the transition from trans-Alfv\'{e}nic to 
super-Alfv\'{e}nic turbulence found by \citet{Heyer+Brunt12}.

\section{Conclusions}

We have argued that MCs are born super-Alfv\'{e}nic with respect to their mean magnetic
field, because of the trans-Alfv\'{e}nic nature of the turbulence in the WISM. We have shown 
that this scenario is supported  by the results of a SN-driven simulation meant to represent an 
overdense region (e.g. a spiral arm) of the ISM on a scale of 250\,pc, with a realistic mean magnetic 
field. MCs selected from the simulation have a mean magnetic-field strength of approximately
10~$\mu$G, only a factor of two  larger than the mean field averaged over the whole computational 
volume. Their internal rms velocity is super-Alfv\'{e}nic with respect to both their mean and rms 
magnetic field strength. 

Despite the small value of their mean magnetic field strength, super-Alfv\'{e}nic MCs are expected
to naturally generate dense cores with stronger magnetic field as the result of compression by 
turbulent shocks. The properties of such cores measured in simulations of super-Alfv\'{e}nic turbulence 
are consistent with those of real MC cores.

\acknowledgements
Computing resources for this work were provided by the NASA High-End Computing (HEC) Program through the NASA Advanced 
Supercomputing (NAS) Division at Ames Research Center. PP acknowledges support by the Spanish MINECO under project AYA2014-57134-P.

\end{document}